# Influencing factors on false positive rates when classifying tumor cell line response to drug treatment

Priyanka Vasanthakumari[1], Thomas Brettin[1], Yitan Zhu[1], Hyunseung Yoo[1], Maulik Shukla[1], Alexander Partin[1], Fangfang Xia[1], Oleksandr Narykov[1], and Rick L. Stevens[1,2]
1 Computing, Environment and Life Sciences, Argonne National Laboratory, Lemont, IL
2 Department of Computer Science, The University of Chicago, Chicago, IL

**Abstract**

Informed selection of drug candidates for laboratory experimentation provides an efficient means of identifying suitable anti-cancer treatments. The advancement of artificial intelligence has led to the development of computational models to predict cancer cell line response to drug treatment. It is important to analyze the false positive rate (FPR) of the models, to increase the number of effective treatments identified and to minimize unnecessary laboratory experimentation. Such analysis will also aid in identifying drugs or cancer types that require more data collection to improve model predictions. This work uses an attention based neural network classification model to identify responsive/non-responsive drug treatments across multiple types of cancer cell lines. Two data filtering techniques have been applied to generate 10 data subsets, including removing samples for which dose response curves are poorly fitted and removing samples whose area under the dose response curve (AUC) values are marginal around 0.5 from the training set. One hundred trials of 10-fold cross-validation analysis is performed to test the model prediction performance on all the data subsets and the subset with the best model prediction performance is selected for further analysis. Several error analysis metrics such as the false positive rate (FPR), and the prediction uncertainty are evaluated, and the results are summarized by cancer type and drug mechanism of action (MoA) category. The FPR of cancer type spans between 0.262 and 0.5189, while that of drug MoA category spans almost the full range of [0, 1]. This study identifies cancer types and drug MoAs with high FPRs. Additional drug screening data of these cancer and drug categories may improve response modeling. Our results also demonstrate that the two data filtering approaches help improve the drug response prediction performance.

**Keywords:** Anti-cancer drug response prediction, data filtering, false positive rate, error analysis

# 1. Introduction

Screening drugs to target specific diseases in the laboratory requires an informed selection of drug candidates. Recent advances in drug discovery such as the high throughput screening facilitates testing a huge number of drugs at a much faster rate than traditional techniques. Cancer is a highly heterogenous disease and precision medicine allows physicians to offer personalized treatment options to patients based on their individual cancer types. Computational methods [1–13] for anti-cancer drug response prediction aids in identifying suitable drugs for specific cancer types based on comprehensive genomic analysis. It is important to analyze the false positive rates (FPR) while using such computational drug response models to minimize the number of experimentation as well as to improve the model performance by collecting more data for experiments with poor FPR.

Our approach in this study is to identify candidate drugs for laboratory screening using a deep neural network classifier (DNN) that predicts whether a cell line will respond to the drug treatment. Several studies[14–17] have generated and reported dose response curves constructed by measuring the growth response of cancer cell lines exposed to drug treatments under study. The dataset used in this work is constructed by including 21 cancer types that have the largest numbers of cell lines with RNA-Seq and drug response data available in a combined set of five drug screening studies. The cell lines are represented by gene expression profiles and the drugs are represented by molecular descriptors. An attention based neural network classification model is used to predict responsive vs. non-responsive treatments. The response labels used for training our classifier are derived by computing the normalized area under the dose response curve (AUC) and assigning a responder or non-responder label to the sample if the AUC is greater or less than some threshold.

This study investigates false positives relative to true positives using the DNN classifier through 100 trials of 10-fold cross-validation. Two data filtering techniques have been applied, including removing samples for which dose response curves are poorly fitted and removing samples whose AUC values are marginal around 0.5 from the training set. This analysis examines the effect of the choice of the threshold that is applied on AUC values for calling response vs. nonresponse, and the selection of samples based on the goodness of fit when computing the dose response curve. Our analysis results show both filtering approaches help improve the model prediction performance. The FPRs of 21 cancer types and 96 different drug mechanism of action (MoA) categories are summarized for the best performing models. The major contributions of this work are: 1) implementing data filtering approaches to improve drug response prediction performance, and 2) identifying cancer types and drug MoAs with high FPRs. Generating more data for cancer types and drug MoA categories with high FPRs, can potentially improve the model performance for predicting responses of these cancer types and drug MoA categories.

# 2. Methods

## 2.1. Data and data splitting strategy

The dataset was constructed by including top 21 cancer types having the greatest number of cell lines with the RNA-Seq and drug response data available in the combined set of CTRP[17], CCLE[14], gCSI[15], NCI-60[18] and GDSC[16] studies. The drug response dataset from the top 21 cancer types consists of 1895 cancer cell lines and 1681 anti-cancer drugs. A sample in the dataset represents a cell line treated with a single drug. The cell lines are represented by gene expression values for 942 benchmark genes from the LINCS[19] study and the drugs are represented by 5270 molecular drug descriptors computed using the Dragon7 software[20].

The AUC value from the dose response curve is calculated and used for determining the classes for the classification task. The experiment samples are categorized into the non-responders (AUC >= 0.5) and responders (AUC < 0.5) and encoded to 0 and 1, respectively. Because drug response data were merged from multiple studies, several experiments (i.e. pairs of cell lines and drugs) had contradicting responses. After removing samples with contradicting responses and duplicated samples, the dataset consisted of 339,896 samples, and the response categories were highly skewed: 95% non-responders, 5% responders. The data is highly unbalanced, as would be expected. There are many more cases when the cancer cells do not respond to the drug treatment.

After building the top21 dataset, we recognized that some experiments did not have a dose response curve fitted well (28% have R-squared fit score < 0.5). Hence the samples with poor R-squared fit scores could confuse model training. Similarly, because of inherent error in fitting the dose-response curve, samples with an AUC close to 0.5 might not be trustworthy. To understand the impacts of samples with a poor dose-response curve fitting and samples near the AUC boundary, we developed three filtering levels (0.0, 0.5, and 0.9) on R-squared fit score and two gap conditions, removing samples around the boundary (AUC between 0.4 and 0.6 - gap1) or samples above the boundary (between 0.5 and 0.7 - gap2). This approach resulted in 10 datasets, each subjected to 100 10-fold cross-validation trials. Table 1 shows the 10 different datasets, their descriptions and the numbers of drugs and cell lines in data.

100 10-fold cross-validation experiments were conducted on each dataset. Each dataset was split into training, validation, and test sets 100 times, with each split being a random 80% training, 10% validation and 10% testing. The samples column in Table 1 shows the number of samples in training, validation, and testing sets. 100 models were trained using the 100 splits of data. The model was trained on the training set and predictions were generated on the testing set for every data split.

## 2.2 Model architecture

This work uses a deep neural network model with attention mechanism. The code base and implementation details of the model (named as 'Attn') can be found in the Benchmarks repository (of the Exa-scale computing project (ECP) application Cancer Distributed Learning Environment (CANDLE)[21]. It utilizes the CANDLE Benchmark infrastructure for quick setup and hyperparameter optimization using the CANDLE Supervisor framework. The Attn model is trained on gene expressions of cancer cell lines and molecular descriptors of drugs to classify the treatment effect as either response or non-response. The trained model is neither cancer nor drug specific.

The model is compiled into 9 layers, consisting of 7 hidden, input and output layers. Input layer has 6212 neurons, the output layer has a single neuron. The 8 hidden layers have 1000, 500, 250, 125, 60, 30, 2. The softmax activation function is used for the first and last hidden layer, while the rest of the layers are set to ReLU. A dropout rate of 0.2 is used. The loss function is set to categorical cross-entropy, which is used for single label prediction, and the SGD optimizer is used.

**Table 1.** Summary of datasets used for 100 trials of 10-fold cross-validation. Each dataset consists of precomputed randomized training, validation, and testing sets. The number of samples in the training, validation, and testing sets, along with the numbers of drugs and cell lines are shown. The table also shows the MCC values obtained after the 100 trials of 10-fold cross-validation were done on each dataset.

| Dataset | # Samples in: Training Validation Testing | Drugs : Cell lines | MCC | Description |
|---|---|---|---|---|
| top21_baseline | 271,915 33,989 33,989 | 1017:1204 | 0.61912237 | Splits of top_21.res_bin.cf_rnaseq.dd_dragon7.labeled.scaled.debug.parquet |
| top21_r.0_baseline | 269,811 33,726 33,726 | 1018:1204 | 0.61809175 | No AUC gap condition. Samples with R-squared fit value less than 0.0 are removed from training. |
| top21_r.0_gap1 | 255,685 33,726 33,726 | 1018:1204 | 0.62442323 | Removes samples from training with the value of the AUC between 0.4 and 0.6, and R-squared fit value less than 0.0 |
| top21_r.0_gap2 | 245,117 33,726 33,726 | 1018:1204 | 0.58580431 | Removes samples from training with the value of the AUC between 0.5 and 0.7, and R-squared fit value less than 0.0 |
| top21_r.5_baseline | 201,895 25,236 25,237 | 991:1204 | 0.61901118 | No AUC gap condition. Samples with R-squared fit value less than 0.5 are removed from training. |
| top21_r.5_gap1 | 187,973 25,236 25,237 | 991:1204 | 0.62321287 | Removes samples from training with the value of the AUC between 0.4 and 0.6, and R-squared fit value less than 0.5 |
| top21_r.5_gap2 | 177,567 25,236 25,237 | 991:1204 | 0.58677235 | Removes samples from training with the value of the AUC between 0.5 and 0.7, and R-squared fit value less than 0.5 |
| top21_r.9_baseline | 144,807 18,101 18,101 | 941:1204 | 0.62977115 | No AUC gap condition. Samples with R-squared fit value less than 0.9 are removed from training. |
| top21_r.9_gap1 | 133,307 18,101 18,101 | 941:1204 | 0.63474431 | Removes samples from training with the value of the AUC between 0.4 and 0.6, and R-squared fit value less than 0.9 |
| top21_r.9_gap2 | 124,565 18,101 18,101 | 941:1204 | 0.59180333 | Removes samples from training with the value of the AUC between 0.5 and 0.7, and R-squared fit value less than 0.9 |

### 2.3 False positive rate analysis

The deep neural network model is trained on each of the 10 datasets described in Table 1. 100 trials of 10-fold cross-validation analysis is performed using each dataset. The performance of the model is analyzed using Mathew's correlation coefficient (MCC). A high value of MCC metric can be obtained only if the model correctly classifies high percentages of both response and nonresponse samples, irrespective of class imbalances. Therefore, due to the high-class imbalance between responders and non-responders, MCC will be a good metric to assess the model classification performance. The effect of the two filtering approaches in building the 10 datasets can be evaluated using the MCC scores. The dataset with the best MCC score is then chosen for further analysis.

The prediction outcomes from the selected dataset are categorized into true positives (TP), true negatives (TN), false positives (FP), and false negatives (FN) by comparing with the ground truth response values. The total numbers of TP, TN, FP, and FN are counted for each cancer type and each drug MoA category separately across all the 100 trials of 10-fold cross-validation. The following evaluation metrics are computed.

$$\text{False positive rate, } FPR = \frac{FP}{(FP+TP)} \tag{1}$$

$$\text{False negative rate, } FNR = \frac{FN}{(FN+TN)} \tag{2}$$

$$\text{Total false rate, } TFR = \frac{(FP+FN)}{(FP+TP+FN+TN)} \tag{3}$$

Because each pair of cell line and drug has been predicted multiple times in the cross-validation analysis, we can also calculate the uncertainty of the prediction using the following equation.

$$\text{Uncertainty} = 1 - \frac{\text{abs}(TP+FP-TN-FN)}{TP+FP+TN+FN} \tag{4}$$

where abs(•) calculates the absolute value of input. The uncertainty measure focuses on the variation of prediction outcome through cross-validation rather than its accuracy. Average uncertainty is first calculated for each cancer cell line, and then it is further averaged across the multiple cell lines of a cancer type. Similarly, average uncertainty is first calculated for each drug and then further averaged across all drugs in a MoA category.

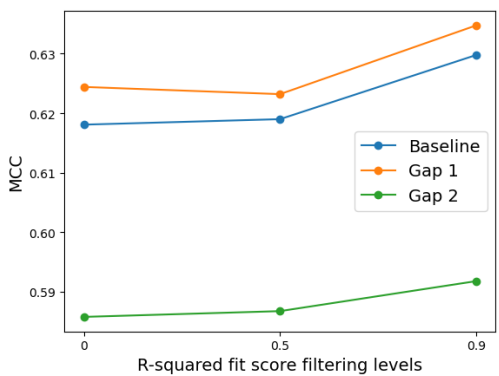

**Figure 1** The variation in MCC with different data filtering approaches and thresholds.

## 3. Results

The MCC column in Table 1 shows the MCC scores obtained when training and evaluating the models on the 10 different datasets. Figure 1 shows the variation in MCC with the three filtering levels on the fitting score of dose response curve, for the three AUC gap conditions, i.e. baseline, gap1 and gap2. The best score is obtained when the dataset is generated using AUC gap1 condition and R-squared fit threshold of 0.9. Therefore, the analysis result of top21_r.9_gap1 dataset is chosen for further analysis. It can also be seen that the models trained on datasets filtered using AUC gap2 condition have much lower performances than the models trained on datasets using baseline or gap1 condition. Another major observation is that filtering using R-squared of 0.9 on dose response curve fitting gives a prediction performance higher than filtering using R-squared of 0 and 0.5 with AUC gap condition unchanged.

The prediction results from the 100 trials of 10-fold cross-validation for models trained on top21_r.9_gap1 dataset are then analyzed. The TP, TN, FP, and FN counted for each cancer type and each drug MoA category are shown in Table 2 and Table 3, respectively. Both Table 2 and Table 3 are sorted according to the false positive rate. Only cancer types with at least 10 cell lines are shown in Table 2. Table 3 includes drug MoA categories with at least one positive sample, i.e. TP + FN > 0. Drug MoA categories without any positive sample, i.e., TP = FN = 0, are not shown in Table 3, because they may indicate "dead" MoA spaces without effective anti-cancer compounds. Average uncertainty for cancer types is shown in Table 2 and that of drug MoA category is shown in Table 3.

Reducing the FPR of predictions is important for improving the cost-efficiency of drug screening practice. A lower FPR leads to more validated effective treatments with the same number of experiments, since usually only experiments predicted to be responsive will be conducted. To reduce the FPR of prediction models, a reasonable strategy is to generate more data of cancer types and drug MoA categories with high FPRs, which are expected to improve the models for predicting responses of these cancer types and MoA categories. As shown in Table 2, the FPR of cancer type ranges from 0.262 to 0.5189. A weighted scheme can be applied to select cancer models from different cancer types for experiments, with the number of cancer models in each cancer type reversely proportional to its FPR. The prediction uncertainties of all cancer types are smaller than 0.05, which are not particularly useful for guiding experiment design. The FPR of drug MoA category spans almost the whole range of [0, 1]. For drug selection, MoA categories with high FPRs can be prioritized. On top of this selection criterion, we can further prioritize MoA categories which also have relatively high FNRs (e.g. > 0.1). Also, there are 12 MoA categories with a prediction uncertainty > 0.1, which can also be prioritized for drug selection.

**Table 2**  Summary of prediction error and uncertainty by cancer type

| Cancer Type | # TN | # TP | # FN | # FP | FNR | FPR | TFR | Uncertainty |
|---|---|---|---|---|---|---|---|---|
| Kidney Renal Clear Cell Carcinoma | 65967 | 890 | 609 | 960 | 0.0091 | 0.5189 | 0.0229 | 0.0233 |
| Ovarian Serous Cystadenocarcinoma | 68210 | 1450 | 654 | 1322 | 0.0095 | 0.4769 | 0.0276 | 0.0276 |
| Colon Adenocarcinoma | 123019 | 3245 | 1563 | 2958 | 0.0125 | 0.4769 | 0.0346 | 0.0254 |
| Breast Invasive Carcinoma | 99705 | 3103 | 1327 | 2349 | 0.0131 | 0.4309 | 0.0345 | 0.0260 |
| Liver Hepatocellular Carcinoma | 45451 | 1273 | 659 | 953 | 0.0143 | 0.4281 | 0.0333 | 0.0233 |
| Lung Squamous Cell Carcinoma | 48626 | 1297 | 519 | 962 | 0.0106 | 0.4259 | 0.0288 | 0.0229 |
| Lung Non-Small Cell Carcinoma | 63813 | 2035 | 881 | 1505 | 0.0136 | 0.4251 | 0.0350 | 0.0207 |
| Lung Adenocarcinoma | 148251 | 4327 | 1994 | 3040 | 0.0133 | 0.4127 | 0.0319 | 0.0276 |
| Lung Small Cell Carcinoma | 79381 | 3361 | 1670 | 2249 | 0.0206 | 0.4009 | 0.0452 | 0.0294 |
| Skin Cutaneous Melanoma | 141321 | 3761 | 1573 | 2434 | 0.0110 | 0.3929 | 0.0269 | 0.0266 |
| Glioblastoma Multiforme | 60407 | 1966 | 800 | 1252 | 0.0131 | 0.3891 | 0.0319 | 0.0281 |
| Stomach Adenocarcinoma | 48636 | 1766 | 1029 | 1075 | 0.0207 | 0.3784 | 0.0401 | 0.0283 |
| Uterine Corpus Endometrial Carcinoma | 55856 | 1971 | 702 | 1174 | 0.0124 | 0.3733 | 0.0314 | 0.0232 |
| Esophageal Carcinoma | 58388 | 2282 | 854 | 1341 | 0.0144 | 0.3701 | 0.0349 | 0.0203 |
| Ovary - other | 42399 | 1187 | 442 | 675 | 0.0103 | 0.3625 | 0.0250 | 0.0201 |
| Pancreatic Adenocarcinoma | 57445 | 2049 | 737 | 1131 | 0.0127 | 0.3557 | 0.0304 | 0.0255 |
| Head and Neck Squamous Cell Carcinoma | 61522 | 2143 | 849 | 1162 | 0.0136 | 0.3516 | 0.0306 | 0.0182 |
| Acute Myeloid Leukemia | 49490 | 3616 | 1943 | 1657 | 0.0378 | 0.3142 | 0.0635 | 0.0444 |
| Lymphoid Leukemia | 137659 | 9521 | 4486 | 3887 | 0.0316 | 0.2899 | 0.0538 | 0.0441 |
| Lymphoid Neoplasm Diffuse Large B-cell Lymphoma | 56624 | 4820 | 2272 | 1757 | 0.0386 | 0.2671 | 0.0615 | 0.0436 |
| Sarcoma | 82321 | 4029 | 1695 | 1430 | 0.0202 | 0.2620 | 0.0349 | 0.0243 |

**Table 3** Summary of prediction error and uncertainty by drug MoA category

| MoA Category | # TN | # TP | # FN | # FP | Uncertainty | FNR | FPR | TFR | #TP + #FN |
|---|---|---|---|---|---|---|---|---|---|
| AKT Inhibitor | 11976 | 0 | 63 | 3 | 0.0333 | 0.0052 | 1.0000 | 0.0055 | 63 |
| ATR Kinase Inhibitor | 2415 | 0 | 5 | 2 | 0.0008 | 0.0021 | 1.0000 | 0.0029 | 5 |
| Bromodomain Inhibitor | 18015 | 0 | 78 | 1 | 0.0007 | 0.0043 | 1.0000 | 0.0044 | 78 |
| Bruton's Tyrosine Kinase (BTK) Inhibitor | 5948 | 0 | 22 | 4 | 0.0036 | 0.0037 | 1.0000 | 0.0044 | 22 |
| FGFR Inhibitor | 18042 | 0 | 14 | 5 | 0.0004 | 0.0008 | 1.0000 | 0.0011 | 14 |
| Hepatocyte Growth Factor Receptor Inhibitor | 3114 | 0 | 9 | 1 | 0.0001 | 0.0029 | 1.0000 | 0.0032 | 9 |
| JNK Inhibitor | 11737 | 0 | 19 | 1 | 0.0002 | 0.0016 | 1.0000 | 0.0017 | 19 |
| Leucine Rich Repeat Kinase Inhibitor | 4957 | 0 | 7 | 26 | 0.0231 | 0.0014 | 1.0000 | 0.0066 | 7 |
| Like (NRF2) Activator | 3431 | 0 | 21 | 6 | 0.0034 | 0.0061 | 1.0000 | 0.0078 | 21 |
| LXR Agonist | 1409 | 0 | 13 | 1 | 0.0019 | 0.0091 | 1.0000 | 0.0098 | 13 |
| MDM Inhibitor | 16860 | 0 | 109 | 5 | 0.0062 | 0.0064 | 1.0000 | 0.0067 | 109 |
| NEDD Activating Enzyme Inhibitor | 5103 | 0 | 83 | 1 | 0.0005 | 0.0160 | 1.0000 | 0.0162 | 83 |
| Nuclear Factor Erythroid Derived | 3431 | 0 | 21 | 6 | 0.0034 | 0.0061 | 1.0000 | 0.0078 | 21 |
| PARP Inhibitor | 10715 | 0 | 98 | 39 | 0.0153 | 0.0091 | 1.0000 | 0.0126 | 98 |
| Pyruvate Dehydrogenase Kinase Inhibitor | 6280 | 0 | 20 | 10 | 0.0023 | 0.0032 | 1.0000 | 0.0048 | 20 |
| RAF Inhibitor | 24327 | 0 | 111 | 46 | 0.0059 | 0.0045 | 1.0000 | 0.0064 | 111 |
| RET Tyrosine Kinase Inhibitor | 23446 | 0 | 60 | 7 | 0.0004 | 0.0026 | 1.0000 | 0.0028 | 60 |
| Retinoid Receptor Agonist | 13625 | 0 | 18 | 2 | 0.0004 | 0.0013 | 1.0000 | 0.0015 | 18 |
| Sphingolipid Biosynthesis Inhibitor | 47 | 0 | 10 | 1 | 0.0357 | 0.1754 | 1.0000 | 0.1895 | 10 |
| XIAP Inhibitor | 4912 | 1 | 73 | 27 | 0.0091 | 0.0146 | 0.9643 | 0.0199 | 74 |
| DNA Synthesis Inhibitor | 13777 | 24 | 398 | 462 | 0.0609 | 0.0281 | 0.9506 | 0.0587 | 422 |
| DNA Alkylating Agent | 12541 | 23 | 296 | 380 | 0.0616 | 0.0231 | 0.9429 | 0.0511 | 319 |
| ALK Tyrosine Kinase Receptor Inhibitor | 12849 | 4 | 98 | 55 | 0.0566 | 0.0076 | 0.9322 | 0.0118 | 102 |
| DNA Methyltransferase Inhibitor | 8398 | 17 | 119 | 99 | 0.0174 | 0.0140 | 0.8534 | 0.0253 | 136 |
| VEGFR Inhibitor | 71888 | 2 | 193 | 9 | 0.0002 | 0.0027 | 0.8182 | 0.0028 | 195 |
| Glucocorticoid Receptor Agonist | 181 | 6 | 36 | 25 | 0.2590 | 0.1659 | 0.8065 | 0.2457 | 42 |
| MCL1 Inhibitor | 8305 | 7 | 76 | 22 | 0.0043 | 0.0091 | 0.7586 | 0.0117 | 83 |
| Ephrin Inhibitor | 5845 | 14 | 300 | 37 | 0.0288 | 0.0488 | 0.7255 | 0.0544 | 314 |
| CHK Inhibitor | 7243 | 37 | 247 | 86 | 0.0099 | 0.0330 | 0.6992 | 0.0437 | 284 |
| KIT Inhibitor | 46879 | 16 | 391 | 36 | 0.0033 | 0.0083 | 0.6923 | 0.0090 | 407 |
| Cell Cycle Inhibitor | 6604 | 56 | 342 | 121 | 0.0383 | 0.0492 | 0.6836 | 0.0650 | 398 |
| Tyrosine Kinase Inhibitor | 8519 | 14 | 287 | 29 | 0.0111 | 0.0326 | 0.6744 | 0.0357 | 301 |
| IGF-1 Inhibitor | 16736 | 13 | 153 | 24 | 0.0017 | 0.0091 | 0.6486 | 0.0105 | 166 |
| MEK Inhibitor | 15268 | 778 | 601 | 1421 | 0.1051 | 0.0379 | 0.6462 | 0.1119 | 1379 |
| ATP Synthase Inhibitor | 198 | 248 | 39 | 405 | 0.4282 | 0.1646 | 0.6202 | 0.4986 | 287 |
| Growth Factor Receptor Inhibitor | 2817 | 32 | 205 | 49 | 0.0194 | 0.0678 | 0.6049 | 0.0819 | 237 |
| Protein Kinase C Inhibitor | 41 | 81 | 23 | 120 | 0.4564 | 0.3594 | 0.5970 | 0.5387 | 104 |
| BCL Inhibitor | 23585 | 158 | 976 | 220 | 0.0326 | 0.0397 | 0.5820 | 0.0480 | 1134 |
| JAK Inhibitor | 26760 | 68 | 615 | 93 | 0.0095 | 0.0225 | 0.5776 | 0.0257 | 683 |
| PDGFR Tyrosine Kinase Receptor Inhibitor | 55829 | 50 | 475 | 62 | 0.0036 | 0.0084 | 0.5536 | 0.0095 | 525 |
| Exportin Antagonist | 6156 | 174 | 261 | 210 | 0.0342 | 0.0407 | 0.5469 | 0.0693 | 435 |
| BCR-ABL Kinase Inhibitor | 18167 | 57 | 432 | 68 | 0.0104 | 0.0232 | 0.5440 | 0.0267 | 489 |
| FLT3 Inhibitor | 39793 | 81 | 454 | 93 | 0.0066 | 0.0113 | 0.5345 | 0.0135 | 535 |
| Inhibitor Of STAT3/JAK2 Signaling | 3805 | 29 | 344 | 33 | 0.0285 | 0.0829 | 0.5323 | 0.0895 | 373 |
| Lipocortin Synthesis Stimulant | 3805 | 29 | 344 | 33 | 0.0285 | 0.0829 | 0.5323 | 0.0895 | 373 |
| STAT Inhibitor | 8382 | 29 | 344 | 33 | 0.0057 | 0.0394 | 0.5323 | 0.0429 | 373 |

| Category | | | | | | | | | |
|---|---|---|---|---|---|---|---|---|---|
| MTOR Inhibitor | 31029 | 1689 | 1659 | 1816 | 0.0520 | 0.0508 | 0.5181 | 0.0960 | 3348 |
| PI3K Inhibitor | 61103 | 1589 | 1133 | 1647 | 0.0233 | 0.0182 | 0.5090 | 0.0425 | 2722 |
| Ribonucleotide Reductase Inhibitor | 10429 | 1157 | 866 | 1172 | 0.1083 | 0.0767 | 0.5032 | 0.1496 | 2023 |
| PKC Inhibitor | 7323 | 2 | 38 | 2 | 0.0004 | 0.0052 | 0.5000 | 0.0054 | 40 |
| EGFR Inhibitor | 48472 | 90 | 641 | 77 | 0.0248 | 0.0131 | 0.4611 | 0.0146 | 731 |
| Dihydrofolate Reductase Inhibitor | 3442 | 290 | 615 | 227 | 0.0549 | 0.1516 | 0.4391 | 0.1841 | 905 |
| Topoisomerase Inhibitor | 19180 | 2952 | 2066 | 2030 | 0.0850 | 0.0972 | 0.4075 | 0.1562 | 5018 |
| CDK Inhibitor | 30761 | 1193 | 821 | 780 | 0.0257 | 0.0260 | 0.3953 | 0.0477 | 2014 |
| Aurora Kinase Inhibitor | 12960 | 25 | 458 | 16 | 0.0062 | 0.0341 | 0.3902 | 0.0352 | 483 |
| NAMPT Inhibitor | 8450 | 2070 | 798 | 1271 | 0.0756 | 0.0863 | 0.3804 | 0.1643 | 2868 |
| HSP Inhibitor | 21623 | 197 | 1060 | 118 | 0.0092 | 0.0467 | 0.3746 | 0.0512 | 1257 |
| Niacinamide Phosphoribosyltransferase Inhibitor | 1452 | 1884 | 476 | 1122 | 0.3576 | 0.2469 | 0.3733 | 0.3239 | 2360 |
| Oxidative Stress Inducer | 18 | 1890 | 8 | 1090 | 0.0162 | 0.3077 | 0.3658 | 0.3653 | 1898 |
| RNA Polymerase Inhibitor | 2362 | 2812 | 165 | 1570 | 0.0571 | 0.0653 | 0.3583 | 0.2511 | 2977 |
| HDAC Inhibitor | 48115 | 5854 | 1128 | 3209 | 0.0321 | 0.0229 | 0.3541 | 0.0744 | 6982 |
| Microtubule Inhibitor | 4491 | 1980 | 140 | 1057 | 0.0910 | 0.0302 | 0.3480 | 0.1561 | 2120 |
| ATPase Inhibitor | 12420 | 2761 | 757 | 1361 | 0.1466 | 0.0574 | 0.3302 | 0.1224 | 3518 |
| Protein Synthesis Inhibitor | 5975 | 536 | 870 | 255 | 0.0805 | 0.1271 | 0.3224 | 0.1473 | 1406 |
| Tubulin Polymerization Inhibitor | 6937 | 11972 | 985 | 5488 | 0.0938 | 0.1243 | 0.3143 | 0.2550 | 12957 |
| PLK Inhibitor | 11877 | 1519 | 1072 | 678 | 0.1384 | 0.0828 | 0.3086 | 0.1155 | 2591 |
| Calcium Channel Blocker | 71 | 2341 | 76 | 795 | 0.0717 | 0.5170 | 0.2535 | 0.2653 | 2417 |
| PKC Activator | 71 | 2341 | 76 | 795 | 0.0717 | 0.5170 | 0.2535 | 0.2653 | 2417 |
| Survivin Inhibitor | 24 | 4145 | 13 | 1396 | 0.0112 | 0.3514 | 0.2519 | 0.2526 | 4158 |
| SRC Inhibitor | 22086 | 254 | 593 | 77 | 0.0132 | 0.0261 | 0.2326 | 0.0291 | 847 |
| Kinesin-Like Spindle Protein Inhibitor | 299 | 562 | 26 | 165 | 0.1188 | 0.0800 | 0.2270 | 0.1815 | 588 |
| Microtubule Stabilizing Agent | 11 | 2956 | 4 | 663 | 0.0113 | 0.2667 | 0.1832 | 0.1835 | 2960 |
| Coflilin Signaling Pathway Activator | 3122 | 818 | 609 | 182 | 0.1179 | 0.1632 | 0.1820 | 0.1672 | 1427 |
| Lim Kinase Activator | 3122 | 818 | 609 | 182 | 0.1179 | 0.1632 | 0.1820 | 0.1672 | 1427 |
| Rock Activator | 3122 | 818 | 609 | 182 | 0.1179 | 0.1632 | 0.1820 | 0.1672 | 1427 |
| Proteasome Inhibitor | 8440 | 4901 | 314 | 1003 | 0.0904 | 0.0359 | 0.1699 | 0.0898 | 5215 |
| NFKB Pathway Inhibitor | 16753 | 4797 | 0 | 628 | 0.0001 | 0.0000 | 0.1158 | 0.0283 | 4797 |
| P21 Activated Kinase Inhibitor | 3552 | 86 | 91 | 8 | 0.0393 | 0.0250 | 0.0851 | 0.0265 | 177 |
| BMX Inhibitor | 4843 | 0 | 7 | 0 | 0.0000 | 0.0014 | | 0.0014 | 7 |
| DNA Dependent Protein Kinase Inhibitor | 6478 | 0 | 19 | 0 | 0.0000 | 0.0029 | | 0.0029 | 19 |
| Farnesyltransferase Inhibitor | 6941 | 0 | 19 | 0 | 0.0000 | 0.0027 | | 0.0027 | 19 |
| G Protein Coupled Receptor Agonist | 1112 | 0 | 7 | 0 | 0.0000 | 0.0063 | | 0.0063 | 7 |
| Glycogen Synthase Kinase Inhibitor | 11360 | 0 | 7 | 0 | 0.0000 | 0.0006 | | 0.0006 | 7 |
| HMGCR Inhibitor | 11336 | 0 | 15 | 0 | 0.0000 | 0.0013 | | 0.0013 | 15 |
| HSP Antagonist | 1284 | 0 | 78 | 0 | 0.0000 | 0.0573 | | 0.0573 | 78 |
| Hypoxia Inducible Factor Inhibitor | 761 | 0 | 41 | 0 | 0.0000 | 0.0511 | | 0.0511 | 41 |
| Ion Channel Antagonist | 3765 | 0 | 8 | 0 | 0.0000 | 0.0021 | | 0.0021 | 8 |
| Map Kinase Inhibitor | 5954 | 0 | 29 | 0 | 0.0000 | 0.0048 | | 0.0048 | 29 |
| Phosphatidylinositol 3-Kinase (Pi3k) Inhibitor | 2310 | 0 | 8 | 0 | 0.0000 | 0.0035 | | 0.0035 | 8 |
| Proteinase Activated Receptor Antagonist | 5097 | 0 | 34 | 0 | 0.0000 | 0.0066 | | 0.0066 | 34 |
| RAD51 Inhibitor | 1294 | 0 | 12 | 0 | 0.0000 | 0.0092 | | 0.0092 | 12 |
| Retinoid Receptor Ligand | 554 | 0 | 7 | 0 | 0.0000 | 0.0125 | | 0.0125 | 7 |
| Ribosomal Protein Inhibitor | 2444 | 0 | 7 | 0 | 0.0000 | 0.0029 | | 0.0029 | 7 |

| Sodium/Hydrogen Exchanger Inhibitor | 1112 | 0 | 7 | 0 | 0.0000 | 0.0063 | | 0.0063 | 7 |
| TP53 Reactivator | 4295 | 0 | 9 | 0 | 0.0000 | 0.0021 | | 0.0021 | 9 |
| Wee1 Kinase Inhibitor | 4068 | 0 | 27 | 0 | 0.0000 | 0.0066 | | 0.0066 | 27 |

## Conclusion

This work performs a comprehensive study with 100 trials of 10-fold cross-validation analysis of anti-cancer drug response data. Two filtering techniques, based on AUC gap condition and the goodness of fitting dose response curve, were used to generate 10 different data subsets. The first part of this work identifies the data subset that gives the best model performance and investigates the effects of the filtering thresholds on the prediction performance. The highest MCC value of 0.635 is achieved in the analysis, which removes samples whose R-squared values resulted from dose response curve fitting are smaller than 0.9 and excludes samples with AUC values in the range of [0.4, 0.6] from the training set. This shows that applying data filtering approaches helps improve the model prediction performance. The second part of the work performs an in-depth analysis on the FPR, TPR and TFR of both cancer types and drug MoA categories, to identify the ones with high FPRs. The FPR of cancer type spans between 0.262 and 0.5189, while that of drug MoA category spans almost the full range of [0, 1]. Collecting more drug screening data for cancer types and drug MoA categories with high FPRs will potentially help developing more accurate prediction models. The post-prediction error analysis conducted in this study can be implemented as a pipeline module to be routinely applied after drug response prediction, which can conveniently provide a summary of prediction errors specific to cancer types and drug categories. This work can also be extended by performing such analysis with several drug response prediction models and conducting a comparison of their results.

## Code availability

The source code of the deep neural network model is available at https://github.com/ECP-CANDLE/Benchmarks/tree/master/Pilot1/Attn